\documentclass{article}
\usepackage{graphicx} 
\usepackage{svg}
\usepackage{amsmath}
\usepackage{pdfpages}
\usepackage{textgreek}
\usepackage[numbers,sort&compress]{natbib}

\title{GPU enabled real-time optical frequency comb spectroscopy and photonic readout}
\author{S.M. Bresler \and D.A. Long \and B.J. Reschovsky \and Y. Bao \and T.W. LeBrun \and J.J. Gorman}
\date{April 2023}

\begin{document}

\maketitle

\section{Abstract}

We describe a GPU-enabled approach for real-time optical frequency comb spectroscopy in which data is recorded, Fourier transformed, normalized, and fit at data rates up to 2.2 GB/s. As an initial demonstration we have applied this approach to rapidly interrogate the motion of an optomechanical accelerometer through the use of an electro-optic frequency comb. However, we note that this approach is readily amenable to both self-heterodyne and dual comb spectrometers for molecular spectroscopy as well as photonic readout where the approach's agility, speed, and simplicity are expected to enable future improvements and applications.
\section{Introduction}

Optical frequency combs are a powerful tool for spectroscopy, metrology, communications, and physical sensing \cite{coddington2016dual, fortier201920, chang2022integrated, lehman2021optical}.  Direct frequency comb spectroscopy allows for broadband spectra to be recorded with high frequency accuracy, coherence, and sampling rates. In many cases a second frequency comb or a continuous wave laser is utilized as a local oscillator to down-convert the optical frequency comb into the radio frequency (RF) domain for digitization without the need for any moving parts \cite{coddington2016dual}. This approach generates large quantities of high frequency data which in turn requires vast storage and computational resources to process.\par

Here we present a solution in which a graphics processing unit (GPU) is utilized as a coprocessor, allowing for essentially infinite acquisition lengths, a two order-of-magnitude reduction in file size, and an increase in computational speed by roughly a factor of 50 when compared to its computer processing unit (CPU) counterpart. In contrast to previously reported methods using field programmable gate arrays (FPGAs) \cite{coddington2010coherent, Roy:12, deschenes2010optical}, the program is readily understood by intermediate scientific programmers, requires comparably small knowledge of the underlying hardware, and leverages the rapid development of GPU-based algorithms intended for use in neural networks, machine vision, and artificial intelligence applications. This approach, which we have denoted \textit{combgpu}, performs data acquisition, signal analysis, and fitting routines in a seamless pipeline. GPU programming allows for high agility, meaning that modifications and new ideas are rapidly testable, and can enable different applications to be run in rapid succession on the GPU; something which is not conducive to FPGA workflows, particularly in complex applications with high data rates, due to compilation and synthesis times, as well as complex hardware description languages which require specialized knowledge and experience.\par

As a demonstration of \textit{combgpu}, we have applied it for the rapid readout of cavity optomechanical accelerometers using an optical frequency comb. Recent work on optomechanical accelerometers has indicated that a comb-based approach to readout has numerous advantages when compared to locking-based techniques \cite{reschovsky2022intrinsically, Long:21}. The most notable of these are increased dynamic range, simplified SI-traceability, and the ability to simultaneously quantify the cavity coupling and mode shape. This application requires a challenging combination of high acquisition rates (near 1 MHz), long acquisition times (exceeding many minutes), and high spectral signal-to-noise ratios \cite{reschovsky2022intrinsically, Long:21}. While we have recently demonstrated approaches which achieve a subset of these targets, the high data volumes, in excess of 2 GB/s, have posed significant obstacles to long-term monitoring, averaging, and feedback. \textit{combgpu} has alleviated these difficulties, allowing for real-time acquisition over essentially indefinite time scales. Further, we note the the computational and experimental methods described herein are readily adapted to a broad range of self-heterodyne or dual comb spectroscopy instruments.\par

\section{Experimental Setup}

\begin{figure}[htbp]
    \centering
    \includegraphics[width = 300pt]{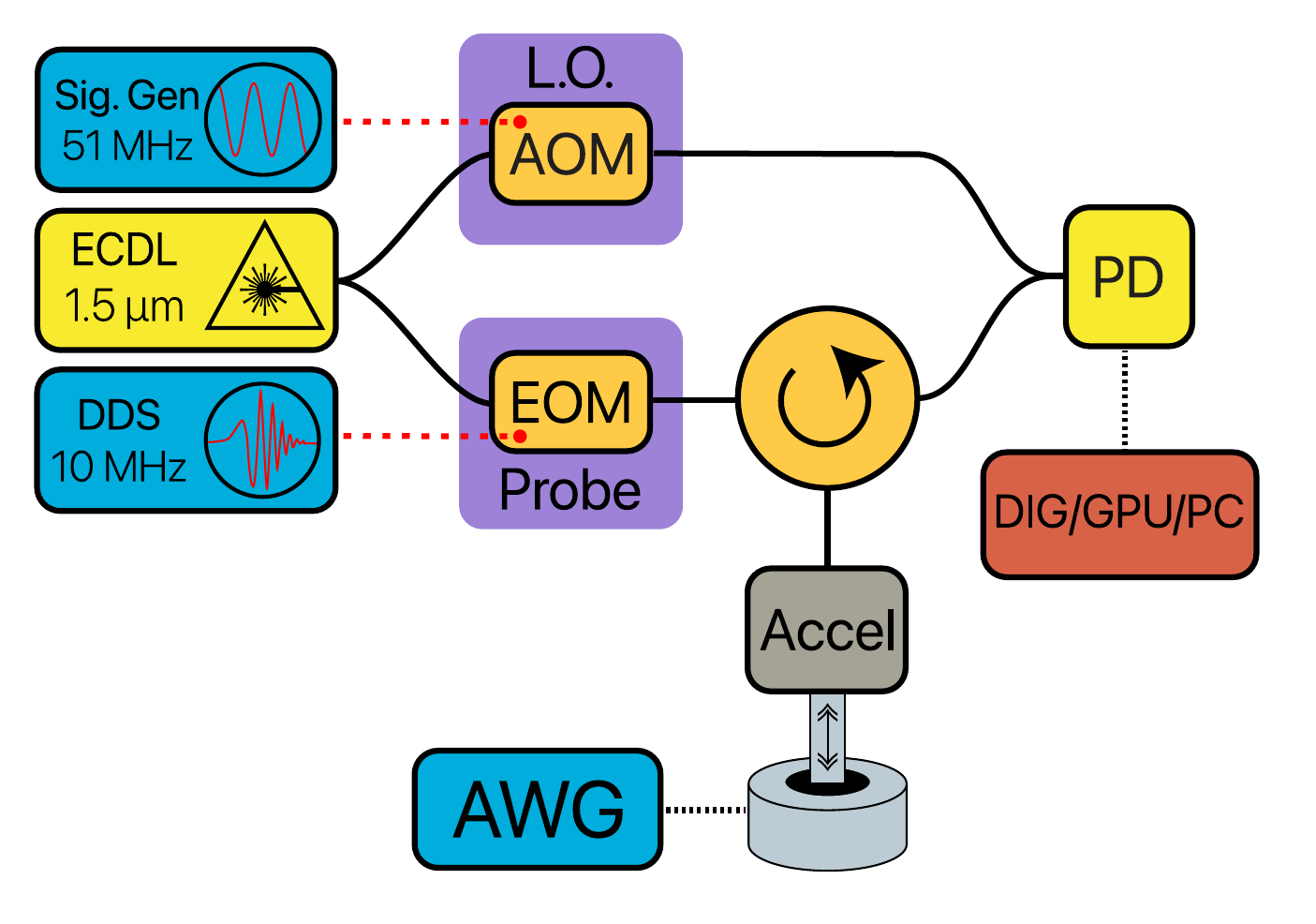}
    \caption{Optical experimental overview. The output of an external-cavity diode laser (ECDL) is split into two paths. The first path is directed through an electro-optic phase modulator (EOM) which is driven by a chirped waveform produced with direct digital synthesis (DDS), producing an optical frequency comb with 10 MHz spacing. The comb then interrogates the accelerometer (Accel) which is placed on an electromechanical shaker table driven by an arbitrary waveform generator (AWG). The second path is frequency shifted by an acousto-optic modulator (AOM) driven by a signal generator (Sig. Gen) to serve as a local oscillator (LO). The two paths are combined on a photodetector (PD) whose output is then sent to the digitizer (DIG) for processing with the graphical processing unit (GPU) within a personal computer (PC).}
    \label{fig:expt-overview}
\end{figure}

The optomechanical accelerometer is composed of a microfabricated stationary concave mirror and a planar mirror attached to a dynamic proof mass suspended by silicon nitride beams, forming an optical cavity with a Fabry-Pérot geometry \cite{zhou2021broadband}. The optical cavity had a finesse of 9200 and a length of 203 μm. The mechanical resonator had a quality factor of 115(1) and a resonance frequency of 23.686(1) kHz. Motion of the proof mass due to an external acceleration gives rise to a displacement of the cavity length and a corresponding shift of a given cavity resonance, which can be readily measured with an optical frequency comb \cite{reschovsky2022intrinsically, Long:21}. Rapidly measuring the cavity mode yields information about the proof mass displacement. The acceleration of the proof mass is then calculable given that the mechanical resonance and quality factor are well known. \ref{fig:expt-overview}  shows the optical layout of the experimental setup which has been described in detail previously \cite{zhou2021broadband, reschovsky2022intrinsically, Long:21}. \par
A self-heterodyne configuration is employed in which an extended cavity diode laser (ECDL) is split into local oscillator (LO) and probe legs. The LO leg shifts the fundamental frequency of the ECDL output by 51 MHz with an acousto-optic modulator (AOM). The probe leg generates a 2.2 GHz wide optical frequency comb with 10 MHz spacing by driving an electro-optic phase modulator (EOM) with a chirped waveform generated via direct digital synthesis \cite{long2019electro}. The optical frequency comb is sent through a circulator to interrogate the optomechanical accelerometer. The light exiting the circulator is then combined with the LO and the resulting RF interferogram is captured with a photodetector and digitized. The accelerometer was placed on an electromechanical shaker table to provide external acceleration. Our implementation of \textit{combgpu}  requires three hardware components in addition to our typical experimental setup: a GPU that is compatible with the Compute Unified Device Architecture (CUDA) platform, a 3.0 GSa/s digitizer with data streaming capabilities,  and a computer that can run Python (version 3.10 or greater) on Windows 10 or greater. \par
The data presented in this work were processed with a 4-core processor clocked at 3.6 GHz and a workstation GPU with 24 GB GDDR6 VRAM and 8192 compute units, configured as a compute device with no graphical rendering responsibilities. The digitizer used in this work had 3 GB of onboard storage and 1.75 GHz of bandwidth.

\section{Data Capture and Processing}

\begin{figure}[htbp]
    \centering
    \includegraphics[width = 300pt]{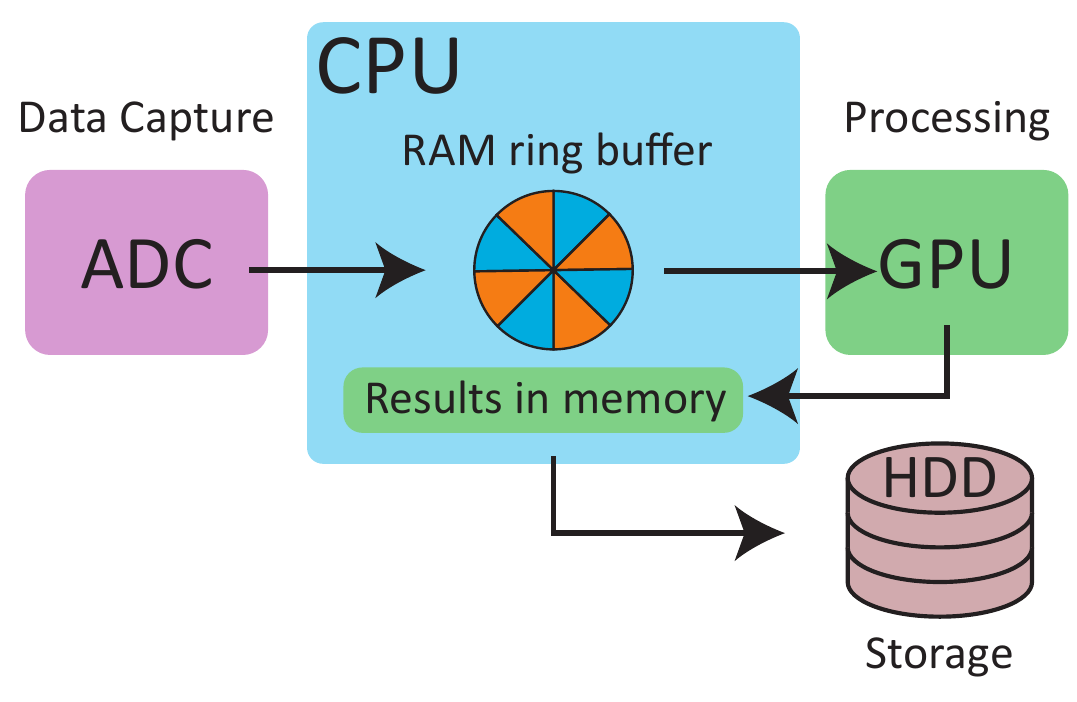}
    \caption{Program function overview. An analog to digital converter (ADC) sends comb data to a ring buffer on the CPU RAM, which is then transferred to the GPU. The GPU performs a mathematical routine on batches of data, which asynchronously complete and are sent back to the CPU. The results are sorted, resulting in a continuous, time ordered series of fit parameters, which are saved to external storage (HDD).}
    \label{fig:overview}
\end{figure}

Figure \ref{fig:overview} shows a high-level overview of \textit{combgpu}. Data from the accelerometer is recorded on the digitizer continuously using a manufacturer supplied field-programmable gate array (FPGA) image. Upon request by the computer, the digitizer initiates a direct memory access routine (DMA), which asynchronously transfers data from the digitizer to a position within a ring buffer on CPU random access memory (RAM). After the CPU receives confirmation that a transfer is complete, the GPU immediately initiates another DMA from RAM to GPU video memory (VRAM). When processing on a portion of the data is complete, the results are transferred back to the CPU’s memory and saved to disk. \textit{combgpu} manages these asynchronous operations across many threads so that data transfer and computation occur in parallel. Indefinite streaming relies on satisfaction of the condition that the rate of data production does not exceed the rate of consumption. If this condition is not met, the buffer eventually overflows with data, and the card ceases to acquire new data until it is reset.\par
\begin{figure}[htbp]
    \centering
    \includegraphics[width = 300pt]{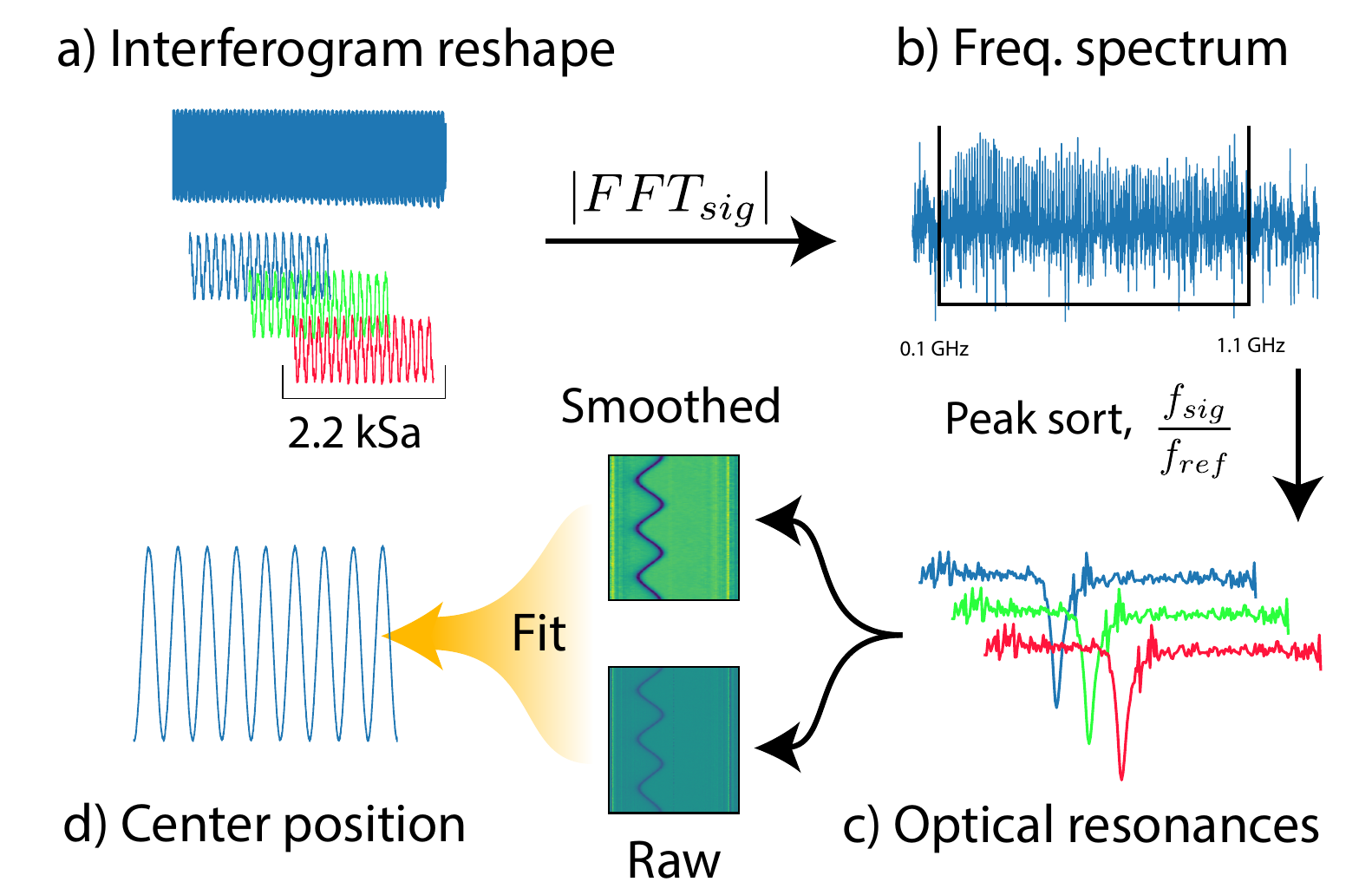}
    \caption{GPU processing overview. (a) A long interferogram is reshaped to form a 2200×N matrix. Each row of this matrix is fast Fourier transformed (FFT) to the frequency domain and compared to a previously calculated reference. (b) The resulting FFT spectrum’s frequencies of interest are then selected to generate (c), which shows an optical resonance of the accelerometer. These optical resonance spectra are averaged spatially and temporally to determine initial parameters and subsequently fit, producing (d), which is transferred back to CPU RAM.}
    \label{fig:processing}
\end{figure}

The processing of raw interferogram data from our accelerometers has several steps which are summarized in Figure \ref{fig:processing}. First, the integer 8-bit data, represented as a 1-dimensional array, must be converted to single precision floating point numbers. This recasting is performed on the GPU to minimize the quantity of data transferred. The new floating point array is then reshaped into a 2-dimensional matrix, with each row containing 2200 temporally sequential samples corresponding to 1 µs of data. A complex-to-real fast Fourier transform (C2R-FFT) is then performed on each row of that array. The bins corresponding to the comb teeth of interest are selected and mapped to reconstruct the frequency comb teeth and therefore the optical spectrum (see Fig \ref{fig:processing}c). Each row is divided by a pre-calculated average reference spectrum, typically recorded at the beginning of the experiment, yielding a group of normalized cavity spectra. Because of the asynchronous nature of the dataflow, batches of cavity spectra might complete out of order, and these are sorted into a temporally contiguous array before saving to non-volatile storage. Overall, the raw data transfer and fitting steps are the most time consuming computational steps, with the FFT step consuming an insignificant portion of the temporal budget.\par

Each optical spectrum in the ensemble can be fit with an arbitrary lineshape to extract dynamic physical information. In our demonstration, we use an asymmetric Fano lineshape \cite{fano1961effects} to represent the cavity resonance, although for other systems more complicated functions (i.e., molecular lineshapes) could be employed. \textit{combgpu} uses \textit{gpufit}, a CUDA accelerated Levenberg-Marquardt fitting program, along with a custom wrapper to expose its functions and CUDA interface in Python \cite{gpufit}. Of the 5 parameters in the Fano lineshape, only the term governing the center position is unconstrained. The two parameters describing the scale and extent of the peak are constrained to about $\pm$10\% of a pre-calculated averaged value. The last two parameters, which govern the asymmetry and width of the peak, are fixed to previously measured values which are device specific. This configuration was used to avoid over-fitting while allowing some tolerance to aberrations in laser power. \textit{combgpu} allows for 20 fit iterations at maximum, with the tolerance in \textit{gpufit} set to $1.0 X 10^{-6}$. \par

As with all non-linear curve fitting processes, convergence on the desired feature and the number of required iterations to reach a specified tolerance is affected by the quality of the initial value, and the center parameter varies strongly under any external stimulation of the accelerometer. Therefore, an initial center value is determined by treating the time-series of the spectra as an image, performing a 2D median filter with a 7x7 footprint on this image, and then selecting the lowest value from each filtered time-series spectrum (see Fig. 3c). Application of this filter avoids an incorrect initial center value, while adding minimal processing time, and importantly reducing the number of iterations required for convergence. With reasonably chosen starting parameters, the typical iterations per converged fit was approximately four.
\par

\section{Experimental Results}

\begin{figure}[htbp]
    \centering
    \includegraphics[width = 400pt]{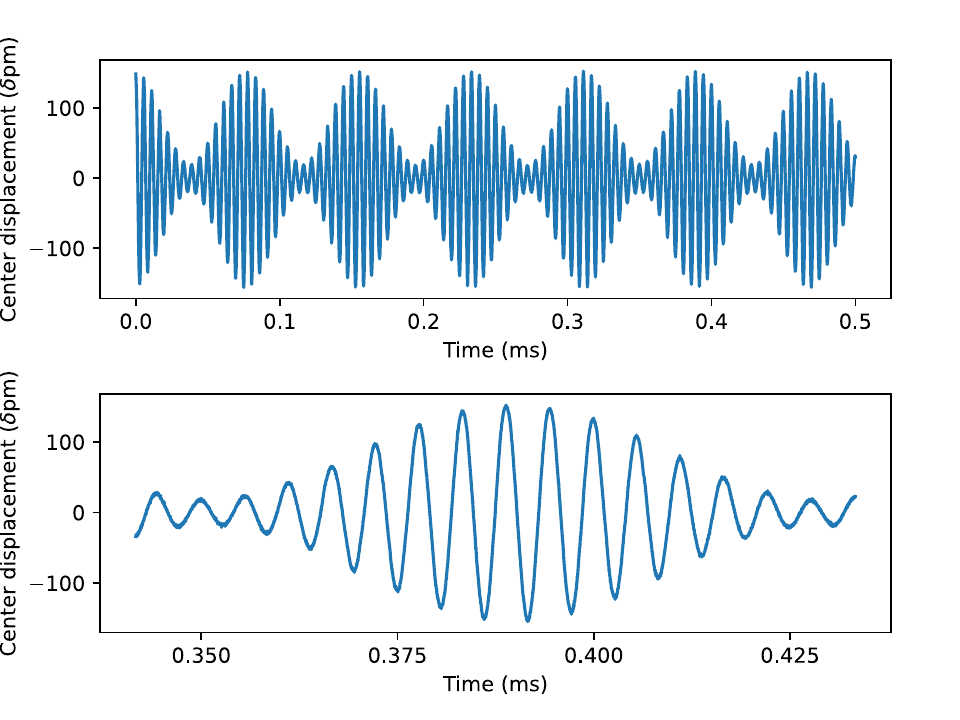}
    \caption{Observed cavity displacement when the electromechanical shaker table was driven by an amplitude modulated sine wave of with a frequency of 3 kHz and a modulation frequency and depth of 213 Hz and 0.80, respectively.  Ten minutes of continuous measurements were recorded with a time resolution of 1 µs with a drive voltage of 6 V. }
    \label{fig:StackedAM}
\end{figure}

As an initial demonstration of this GPU-based approach, we measured the optomechanical cavity displacement when the electromechanical shaker was driven with an amplitude modulated sine wave (see Fig. \ref{fig:StackedAM}). Large cavity displacements, well in excess of a cavity line width, can be readily quantified (something which is difficult with many locking-based methods) \cite{Long:21}. The amplitude modulation frequency of 213 Hz was selected to demonstrate the continuity of each processed segment in the final fit results given the asynchronous nature of the dataflow. We note that one to ten minutes of continuous data was collected for each drive voltage during this measurement with only a portion of those acquisitions shown here.

\begin{figure}[htbp]
    \centering
    \includegraphics[width = 400pt]{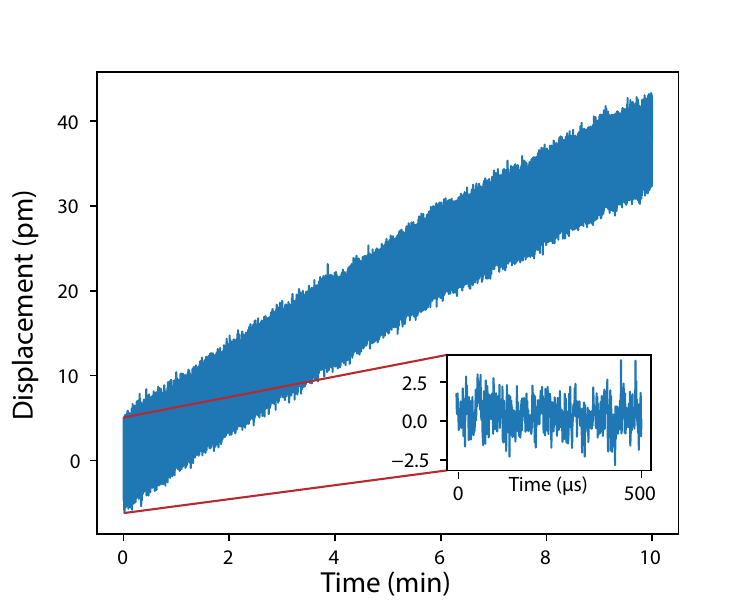}
    \caption{Cavity displacement extracted from a 10 minute subset of continuously recorded data when the electromechanical shaker was disabled. The inset shows a 500 µs long subset of the acquisition. The drift of the cavity length is attributed to temperature and humidity changes during the acquisition. We note that these types of measurements were previously limited to 0.5 s records due to the on-board memory of the digitizer card \cite{reschovsky2022intrinsically, Long:21}.}
    \label{fig:still-long}
\end{figure}

\begin{figure}[htbp]
    \centering
    \includegraphics[width=350pt]{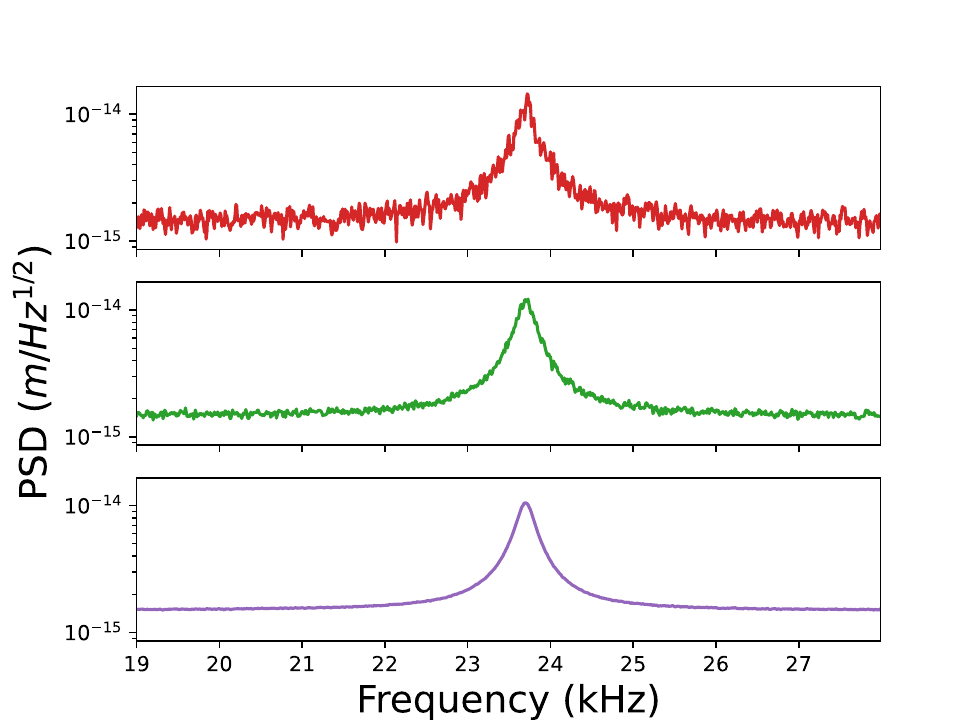}
    \caption{Power spectral density of portions of the accelerometer displacement data found in Figure 4. The fundamental mechanical resonance of the optomechanical accelerometer is visible near 24 kHz. The effect of increasing the duration of data on the fidelity of the recovered resonance is shown by comparing times of 1 s (top, red), 10 s (middle, green), and 10 min (bottom, purple).A fit to the top trace yields a mechanical resonance of  f= 23.706(1) kHz and a mechanical quality factor of Q = 120(1). A fit to the bottom trace yields f = 23.699827(82) kHz and Q = 114.4(1)}
    \label{fig:FFTStill}
\end{figure}

To examine the noise performance of the generation and readout method combined with the accelerometer, we recorded a ten-minute record when the shaker was not excited (see Figure \ref{fig:still-long}). Displacement of the cavity length is observable over long timescales due to thermal and humidity drifts. Figure \ref{fig:FFTStill} shows the corresponding power spectral density for various averaging times.  This measurement of the accelerometer’s thermomechanical resonance is crucial for accurate accelerometry because the mechanical resonance frequency and quality factor are the two parameters needed to invert the measured cavity displacement into acceleration \cite{zhou2021broadband, reschovsky2022intrinsically, gabrielson1993mechanical}. Here the ability to perform long-time averaging can play an important role in reducing the uncertainty associated with these parameters, leading to a reduction of the overall acceleration uncertainty.

The CPU-based data processing we previously employed was subject to several technical limitations. First, the absence of an FPGA streaming image meant that the maximum duration of any single recording was limited to the 3 GB capacity of the onboard digitizer RAM module. In our use-case, recording at 2.2 GSa/s with 8-bit data, this corresponded to a maximum duration of about 1.4 s. Second, processing that length of data took approximately 30 seconds, yielding an overall data throughput rate of 1.7\%. Finally, 1.4 s of that data had a file size of approximately 3 GB. \textit{combgpu}, in contrast, has calculated and recorded accelerometer positions continuously for 40 minutes, and those results are immediately available and continuously updated to a file that is approximately 30 GB. A theoretical CPU-only implementation of \textit{combgpu} would take 15 hours to process this duration of data, and the continuous raw interferogram would have a file size of 5.3 TB. Importantly, the hardware required to achieve these numbers in the CPU-only version involved a 24-core workstation processor which is comparable to the cost of the GPU used in this demonstration.

\section{Discussion}

While \textit{combgpu} has dramatically improved capabilities in comparison to existing methods, it still has several limitations and inefficiencies that can be improved. First, the CPU ring buffer is currently acting as a transfer hub for the digitizer, CPU, and GPU memory. Remote Direct Memory Access (RDMA) is a technology that allows for the digitizer card to bypass CPU RAM entirely to write directly to GPU global memory \cite{glaser2015strong}. Further, the bandwidth of the peripheral component interface express (PCI-e) 3.0 interface which manages traffic between the CPU and GPU can be doubled by a using PCI-e 4.0 bus. If all components saturated 16 PCI-e 4.0 lanes and RDMA were implemented, the total data transfer time would decrease by a factor of about 4. Next, the digitizer is underclocked from its nominal clock rate of 3 GHz to 2.2 GHz, which limits the resolvable span of the comb by about 0.4 GHz. It is notable that any potential aliasing caused by the Nyquist frequency of the sampling being close to the comb span was not detected, which is attributed to any aliased frequency component being nonlocal to the comb teeth selected for processing. The digitizer also operates at 8-bit resolution rather than its maximal resolution of 12-bits to reduce transfer time. Further, the latency of the entire data pipeline for any frame of data is approximately 100 ms as configured for this experiment. As transfer times improve, full resolution data and greater sampling frequencies at lower latency become tractable. The current iteration of the program has enough temporal overhead to perform 20 iterations of fits per batch of 60,000 fits, which  allows for a convergence tolerance of 1 in 10,000 in typical conditions. Faster transfer times would improve this number. Finally, the program could be rewritten to leverage Open Computing Language (OpenCL) and other GPU-centric languages that would enable compatibility with a wide variety of GPUs.

\textit{combgpu} fundamentally advances our optomechanical accelerometer technology from being a proactive measurement, where excitations are carefully synchronized and pre-engineered, to a device that can passively monitor acceleration. Seismic activity detection is a particularly motivating use-case, being a largely stochastic event that is difficult to anticipate. The output data from the processing could also be subjected to further meta-analysis with a more efficient transfer process. Further improvements to latency also open opportunities for a closed-loop feedback system. For example, the accelerometer’s  motion could be counteracted by driving a stage it is resting on to accelerate in the opposite direction. Finally, we note that the GPU processing approach described herein is adaptable to atomic and molecular spectroscopy, which would allow for real time sensing and deep averaging in both self-heterodyne and dual comb systems while readily managing the large data volumes. Further, the data scheme shown here offers a blueprint for any high data-rate experiment that desires streaming with post-processing. We anticipate that these applications can greatly benefit from the approach described herein and that these benefits will only magnify as GPU technology continues to rapidly advance.

\section{Acknowledgements}
Certain equipment, instruments, software, or materials are identified in this paper in order to specify the experimental procedure adequately.  Such identification is not intended to imply recommendation or endorsement of any product or service by NIST, nor is it intended to imply that the materials or equipment identified are necessarily the best available for the purpose. This research was performed in part in the NIST Center for Nanoscale Science and Technology Nanofab and supported by the NIST-on-a-Chip program.

\bibliographystyle{unsrtnat}
\bibliography{main.bib}

\end{document}